\documentstyle[13lomcon,cite]{article}
\begin{document}

\newcommand{\be}{\begin{equation}}
\newcommand{\ee}{\end{equation}}
\newcommand{\ds}{\displaystyle}
\newcommand{\vp}{\varphi}
\newcommand{\vx}{\vec{x}}
\newcommand{\vy}{\vec{y}}
\newcommand{\vz}{\vec{z}}
\newcommand{\vk}{\vec{k}}
\newcommand{\vq}{\vec{q}}
\newcommand{\vpp}{\vec{p}}
\newcommand{\vn}{\vec{n}}
\newcommand{\vg}{\vec{\gamma}}

\title{CHIRAL SYMMETRY BREAKING AND THE LORENTZ NATURE OF CONFINEMENT}

\author{A.V.Nefediev\footnote{e-mail: nefediev@itep.ru}}

\address{Institute of Theoretical and Experimental Physics, B.Cheremushkinskaya 25, 117218 Moscow, Russia}

\maketitle\abstracts{The Lorentz nature of confinement in a heavy--light quarkonium is investigated. 
It is demonstrated that an effective scalar interaction is generated selfconsistently as a result of chiral
symmetry breaking, and this effective scalar interaction is responsible for the QCD string formation between
the quark and the antiquark.}

The question of the Lorentz nature of confinement is one of the most long-standing problems in QCD. Indeed, this question is very important
for understanding the phenomenon of chiral symmetry breaking, for establishing the correct form of spin-dependent potentials in heavy
quarkonia, for understanding the relation between chiral symmetry breaking and the QCD string formation between quarks, and so on. The latter question
is discussed in this talk at the example of a heavy--light quarkonium. 

The approach of the QCD string \cite{DKS} appears very convenient and successful
in studies of various properties of hadrons, both conventional and exotic. This approach follows naturally
from the Vacuum Background Correlators Method (VCM) \cite{VCM}. 
The key idea of the approach is description of the gluonic degrees
of freedom in hadrons in terms of an extended object --- the QCD string --- formed between colour sources. Nonexcited string is approximated
by the straight--line ansatz, so that only radial scratchings and rotations are allowed, 
whereas excitations of the string are described by
adding constituent gluons to the system, so that a hybrid meson, for example, can be represented as the quark--antiquark pair attached to the
gluon by two straight--line segments of the string. 

Consider the simplest case of the quark--antiquark conventional meson consisting of a
quark and an antiquark connected by the straight--line Nambu--Goto string with the tension $\sigma$. The Lagrangian of such a system is
\be
L=-m_1\sqrt{1-\dot{\vx}_1^2}-m_2\sqrt{1-\dot{\vx}_2^2}
-\sigma r\int_0^1d\beta\sqrt{1-[\vec{n}\times(\beta\dot{\vec{x}}_1+(1-\beta)\dot{\vec{x}}_2)]^2},
\label{L}
\ee
$$
\vec{r}=\vx_1-\vx_2\quad \vec{n}=\vec{r}/r,
$$
and one can proceed along the lines of Ref.~\cite{DKS} in order to arrive at the Hamiltonian. An important ingredient which
distinguishes the given approach from the naive potential approach is the proper dynamics of the string which is encoded in the
velocity--dependent interaction (the last term in Eq.~(\ref{L})) and which translates into an extra inertia of the system with respect to the
rotations. This extra inertia leads to the decrease of the (inverse) Regge trajectories slope and brings it to the experimentally observed
value of $2\pi\sigma$ or $\pi\sigma$, for a light--light and heavy--light systems, respectively \cite{DKS,MNS}. This effect is unimportant for our present
purposes and thus we neglect it, arriving a simple Salpeter equation (which is exact for the case of
vanishing angular momentum, $l=0$):
\be
H\psi=M\psi,\quad H=\sqrt{\vpp^2+m_1^2}+\sqrt{\vpp^2+m_2^2}+\sigma r,
\label{Hll}
\ee
which is celebrated in the literature. The question we address in this talk is whether the confining interaction in Eq.~(\ref{Hll}) is of the
scalar of vectorial nature. In order to make things as simple as possible, we consider the one--particle limit of Eq.~(\ref{Hll}) setting
$m_1\to\infty$ and $m_2\equiv m$. In this case, the resulting effectively single--particle system can be described by a Dirac-like equation
and the question posed above translates into the form of the Dirac operator in this equation, that is whether the confining interaction is
added to the energy or to the mass term in this operator. 

We choose the following strategy \cite{hlya}. We start from the Euclidean Green's function
of the given heavy--light quarkonium,
$$
S_{q\bar Q}(x,y)=\frac{1}{N_C}\int D{\psi}D{\psi^\dagger}DA_{\mu}\;
\psi^\dagger(x) S_{\bar Q} (x,y|A)\psi(y)
$$
\be
\times\exp{\left\{-\frac14\int d^4x F_{\mu\nu}^{a2}-\int d^4x
\psi^\dagger(-i\hat \partial -im -\hat A)\psi \right\}},
\label{Sqq}
\ee
and fix the modified Fock--Schwinger gauge \cite{FSg},
\be
\vec{x}\vec{A}(x_4,\vec{x})=0\quad A_4(x_4,\vec{0})=0,
\ee
which leads to the static particle decoupling from the system. Then we perform integration over the gluonic field in Eq.~(\ref{Sqq}) and
arrive at the Dyson--Schwinger-type equation derived in the Gaussian approximation for the QCD vacuum \cite{VCM}, that is only the bilocal
correlator of gluonic fields is retained:
\be
(-i\hat{\partial}_x-im)S(x,y)-i\int d^4zM(x,z)S(z,y)=\delta^{(4)}(x-y),
\label{DS}
\ee
with the mass operator $-iM(x,z)=K_{\mu\nu}(x,z)\gamma_{\mu}S(x,z)\gamma_{\nu}\vphantom{\int}$.
Following Ref.~\cite{NS} we approximate the interaction kernel as
\be
K_{44}(x,y)\equiv K(x,y)\approx \sigma(|\vx|+|\vy|-|\vx-\vy|),\; K_{4i}(x,y)=0,\; K_{ik}(x,y)=0
\label{kernel}
\ee
and rewrite Eq.~(\ref{DS}) in the form of a Schr{\" o}dinger-type equation (in Minkowskii space):
\be
(\vec{\alpha}\hat{\vpp}+\beta m)\Psi(\vx)+\beta\int d^3z M(\vx,\vz)\Psi(\vz)=E\Psi(\vx),
\label{Seq}
\ee
with the mass operator
\be
M(\vx,\vz)=-\frac{i}{2}K(\vx,\vz)\beta\Lambda(\vx,\vz),\quad \Lambda(\vx,\vz)=2i\int\frac{d\omega}{2\pi}S(\omega,\vx,\vy)\beta.
\ee
The question of the Lorentz nature of confinement can be now formulated as the question of the matrix structure of the quantity $\Lambda$
\cite{hlya}. Indeed, the phenomenon of spontaneous breaking of chiral symmetry (SBCS) means that a piece 
proportional to the matrix $\beta$ appears in $\Lambda$ {\em in a
selfconsistent way}. A convenient technique to deal with the phenomenon of chiral symmetry breaking is given by the chiral angle approach \cite{qm}.
Let us parametrise the quark selfinteraction, described by the term $V(\vx-\vy)=\sigma|\vx-\vy|$ in the kernel (\ref{kernel}), 
\be
\Sigma(\vec{p})=\int\frac{d^4k}{(2\pi)^4}\gamma_0\frac{-iV(\vec{p}-\vec{k})}{\gamma_0 k_0-\vg\vk-m-\Sigma(\vk)}\gamma_0=
[A_p-m]+(\vec{\gamma}\hat{\vec{p}})[B_p-p],
\label{sgma}
\ee
by means of the so-called chiral angle $\vp_p$. Then the selfconsistency condition of such a
parametrisation, $A_p/B_p=\tan\varphi_p$, takes the form (explicit expressions for the $A_p$ and $B_p$ follow
from the term by term comparison of the l.h.s. and the r.h.s. of Eq.~(\ref{sgma})):
\be
ps_p-mc_p=\frac{1}{2}\int\frac{d^3k}{(2\pi)^3}V(\vpp-\vk)\left[c_ks_p-(\hat{\vec{p}}\hat{\vec{k}})s_kc_p\right],
\;s_p/c_p=\sin/\cos\vp_p,
\ee
which is known as the mass--gap equation \cite{qm}. Nontrivial solution to this equation behaves as 
$\vp_p(p=0)=\pi/2$ and $\vp_p(p\to\infty)=0$. This allows us to make definite conclusions concerning the 
Lorentz nature of confinement in Eq.~(\ref{Seq}) depending on the value taken by the quark relative momentum. Indeed,
\be
\Lambda(\vpp,\vq)=2i\int\frac{d\omega}{2\pi}S(\omega,\vpp,\vq)\beta=(2\pi)^3\delta^{(3)}(\vpp-\vq)[
\beta\sin\vp_p-(\vec{\alpha}\hat{\vpp})\cos\vp_p],
\ee
and thus
\be
\Lambda(\vpp,\vq)\propto
\left\{
\begin{array}{llll}
\beta,&p\to 0&\Longrightarrow&{\rm scalar\;confinement}\\
(\vec{\alpha}\vpp),&p\to\infty&\Longrightarrow&{\rm vectorial\;confinement.}
\end{array}
\right.
\label{lmts}
\ee
In the meantime, Eq.~(\ref{Seq}) admits a Foldy--Wouthuysen transformation, which can be performed in a closed
form \cite{KNR},
\be
\Psi(\vec{p})=T_p{\psi(\vec{p})\choose 0},\quad
T_p=\exp{\left[-\frac12(\vec{\gamma}\hat{\vec{p}})\left(\frac{\pi}{2}-\vp_p\right)\right]}.
\ee
The resulting equation reads:
\be
E_p\psi(\vec{p})+\int\frac{d^3k}{(2\pi)^3}V(\vpp-\vk)\left[C_pC_k+
({\vec \sigma}\hat{\vpp})({\vec
\sigma}\hat{\vk})S_pS_k\right]\psi(\vec{k})=E\psi(\vec{p}),
\label{Seq2}
\ee
where $C_p/S_p=\cos/\sin\frac12\left(\frac{\pi}{2}-\vp_p\right)$
and $E_p=A_p\sin\vp_p+B_p\cos\vp_p$ is the dressed--quark dispersive law which can be reasonably well 
approximated by the free-quark dispersion (this approximation fails for the chiral pion --- see, for
example, the discussion in Ref.~\cite{NR}).
For small values of the interquark momentum, when $\vp_p\approx\pi/2$, Eq.~(\ref{Seq2}) reduces to the
one--particle limit of the Salpeter Eq.~(\ref{Hll}). Notice that, according to Eq.~(\ref{lmts}), this limit exactly corresponds to the purely
scalar confinement in Eq.~(\ref{Seq}). We conclude therefore that the intrinsic Lorentz nature of the QCD
string is scalar. As soon as chiral symmetry is broken spontaneously, the selfconsistently generated scalar 
part appear in the effective interquark interaction. If this interaction dominates, the system can be
described by the Salpeter Eq.~(\ref{Hll}) or by its more sophisticated version which takes the proper string
dynamics into account.

\section*{Acknowledgments}

This work was supported by the Federal Agency for Atomic Energy of Russian 
Federation and by grants NSh-843.2006.2, DFG-436 RUS 113/820/0-1(R), 
RFFI-05-02-04012-NNIOa, and PTDC/FIS/70843/2006-Fi\-si\-ca.

\end{document}